# COLLABORATIVE BUSINESS INTELLIGENCE VIRTUAL ASSISTANT

Olga Cherednichenko and Fahad Muhammad

*Univ Lyon, Univ_Lyon 2, UR ERIC – 5 avenue Mendès France, 69676 Bron Cedex, France*

**Abstract**
The present-day business landscape necessitates novel methodologies that integrate intelligent technologies and tools capable of swiftly providing precise and dependable information for decision-making purposes. Contemporary society is characterized by vast amounts of accumulated data across various domains, which hold considerable potential for informing and guiding decision-making processes. However, these data are typically collected and stored by disparate and unrelated software systems, stored in diverse formats, and offer varying levels of accessibility and security. To address the challenges associated with processing such large volumes of data, organizations often rely on data analysts. Nonetheless, a significant hurdle in harnessing the benefits of accumulated data lies in the lack of direct communication between technical specialists, decision-makers, and business process analysts. To overcome this issue, the application of collaborative business intelligence (CBI) emerges as a viable solution. This research focuses on the applications of data mining and aims to model CBI processes within distributed virtual teams through the interaction of users and a CBI Virtual Assistant. The proposed virtual assistant for CBI endeavors to enhance data exploration accessibility for a wider range of users and streamline the time and effort required for data analysis. The key contributions of this study encompass: 1) a reference model representing collaborative BI, inspired by linguistic theory; 2) an approach that enables the transformation of user queries into executable commands, thereby facilitating their utilization within data exploration software; and 3) the primary workflow of a conversational agent designed for data analytics.

**Keywords** artificial intelligence, collaborative business intelligence, virtual assistance, machine learning, model

## 1. Introduction

The decision-making process is complex and, as a rule, depends significantly on the information that the person who makes the decision owns. Today's world is characterized by huge volumes of accumulated data in various domains. These data can be really helpful in terms of preparing and making the decisions. However, this data are collected and stored by different unrelated software systems, stored in different formats, providing different levels of access and security. In addition, these data may be incomplete, contradictory, unreliable. To solve problems associated with the processing of large volumes of data, they turn to data analysts. Business intelligence (BI) helps you gain valuable insights and make strategic decisions. Business intelligence tools analyze historical and current data and present the results in intuitive visual formats. A significant obstacle to achieving the effect of using the accumulated data is the lack of direct communication between technical specialists and decision makers and business process analysts. The solution to this problem is to apply the approach of collaborative business intelligence (CBI).

As the analysis shows, the need for data research arises not only from business in order to increase its profits, or the government to solve national problems, but also from society and individual citizens to understand and justify socially significant or private decision-making. In this case, it is quite difficult to organize the interaction of potential users, decision makers, and technical specialists in data analysis. The project BI4people [1] is created to help solve these problems. The aim of BI4people is to bring the power of Business Intelligence to the largest possible audience, by implementing the data warehousing process in software-as-a-service mode, from multisource, heterogeneous data integration to intuitive analysis and data visualization [1,2]. The main idea is to collect people in one virtual and encourage them leave their comment or opinions for general purpose. Moreover, reusing another collaborators' results or comments makes general BI - Collaborative BI.

Building upon the current cutting-edge trends, this paper extends our previous research [3] by exploring the potential of conversational chatbots in enhancing CBI tools and introducing novel opportunities for non-experienced users. In our prior work [3], we proposed a reference model for a virtual collaborative assistant in CBI, comprising three key components: conversational, data exploration, and recommendation agents. We contend that adopting a comprehensive approach based on these three elements forms the foundation for a virtual collaboration tool. The present study highlights the role of a conversational agent in assisting data exploration analysis and investigates the essential features that can facilitate the transformation of user queries into executed commands, particularly for non-experienced users. Furthermore, this research focuses on addressing the interface challenges between the conversational and data exploration components.

## 2. Background

Data exploration is an important component of the BI process, which involves collecting, identifying, and analyzing data to discover meaningful insights and patterns. The main goal of data exploration is to identify key business opportunities and challenges that can drive decision-making and improve business performance.

The data exploration process involves systematically investigating and analyzing datasets to gain insights, identify patterns, and uncover meaningful information. It is a crucial step in data analysis and, in general, the data exploration process consists of the following steps:
1. Data collection: Gathering relevant data from various sources, such as databases, files, or APIs.
2. Data cleaning and preprocessing: Removing inconsistencies, errors, and missing values from the data. This step may also involve transforming the data into a suitable format for analysis.
3. Descriptive statistics: Calculating summary statistics, such as mean, median, standard deviation, and generating visualizations (e.g., histograms, scatter plots) to understand the distribution and basic characteristics of the data.
4. Data visualization: Creating visual representations (e.g., charts, graphs, heatmaps) to explore patterns, relationships, and trends within the data.
5. Exploratory data analysis: Conducting in-depth analysis by applying statistical techniques, data mining algorithms, or machine learning methods to uncover hidden patterns, correlations, or anomalies in the data.
6. Hypothesis testing: Formulating and testing hypotheses to validate or refute assumptions about the data and draw meaningful conclusions.
7. Iterative refinement: Iteratively repeating the above steps, refining analysis techniques, and exploring different perspectives to gain deeper insights and enhance understanding of the data.

The data exploration process is iterative and interactive, involving a combination of manual exploration and automated analysis techniques. Its goal is to reveal valuable information, generate hypotheses, and provide a foundation for further analysis and decision-making processes. The most formidable aspect lies within formulating assumptions about the data and deriving significant conclusions. Conversely, users may possess inquiries pertaining to the domain that can be addressed through data analysis. However, such users must possess expertise in data analysis, as well as proficiency in the relevant tools and methodologies, in order to effectively manipulate and explore the data.

The current state of collaborative business intelligence can be characterized as an expanding field that continually witnesses the development of numerous tools and techniques, all aimed at facilitating effective collaboration within decision-making processes for teams. Noteworthy advancements within the CBI domain encompass the incorporation of social media functionalities, enhanced accessibility through mobile devices, and the adoption of cloud-based solutions. These innovations have empowered users to engage in collaborative work and access data from any location, utilizing any device, and at their preferred time. Furthermore, the integration of natural language processing (NLP) and machine learning (ML) has simplified user interactions with data, enabling the extraction of valuable insights and enhancing the efficiency and effectiveness of decision-making processes. An additional benefit is the utilization of natural language querying (NLQ), which facilitates non-technical users' access to and analysis of data, thereby streamlining their involvement in the decision-making process.

The utilization of robotic assistants to aid users has become progressively prevalent in e-commerce platforms and web portals with diverse functionalities. The present study aims to explore the potential advantages offered by conversational chatbots in enhancing the data exploration process and introducing novel prospects within CBI. A chatbot refers to a computer program designed to simulate human conversation, engaging in text or voice-based interactions with users [4]. Chatbots offer several notable benefits, including round-the-clock availability, expedited response times, personalized interactions, scalability, and cost-effectiveness.

The core objective of our research revolves around developing a model for CBI processes within distributed virtual teams, achieved through the interaction between users and a CBI Virtual Assistant [3]. The primary aim of the virtual assistant is to enhance the accessibility of data exploration for a broader user base while simultaneously reducing the time and effort required for data analysis. This research endeavors to introduce innovative conversational agents that transform the user's interaction with data and ML models, ultimately making data science more accessible to a wider range of users.

Consequently, the following research questions are posed:

RQ1. Can the implementation of a virtual assistant utilizing a conversational interface effectively simplify the data exploration process for non-experts?

RQ2. Which specific features of a virtual assistant hold the utmost significance in assisting users with limited knowledge and expertise?

## 3. The state of the art

Categorization of chatbots is based on their functional capabilities and the nature of collaboration they enable [5, 6, 7]. These intelligent systems are equipped to provide valuable insights, recommendations, and predictions by harnessing the power of available data. Furthermore, chatbots have the ability to send notifications and alerts to users in response to predetermined triggers, such as alterations in data or anomalies detected in critical metrics.

However, it is important to note that while chatbots fall under the umbrella term of conversational agents (CAs), not all CAs are chatbots. CAs encompass a broader range of computer programs or systems that engage in natural language interactions with users [8, 9]. These agents can be rule-based or employ ML and NLP techniques to understand and respond to user inputs.

An extensive review of 233,085 papers conducted by the authors of [10] revealed that only 81 papers met the inclusion criteria, which considered factors such as relevant abstracts, clear presentation of methodology, availability of full-text articles, relevance, and usage of the English language. The findings from [10] indicate that the most commonly co-occurring keywords in the selected papers are "chatbot" and "artificial intelligence". It is also noteworthy that the Python programming language is widely employed in chatbot development [10]. These observations lead to the conclusion that while the topic is not novel, it remains cutting-edge, with numerous successful implementations of chatbots and conversational agents demonstrating their potential. Additionally, a variety of tools and language models are available for implementing personal assistants.

In order to construct contextual knowledge within a dialogue, several factors need to be considered, such as retaining the conversation history, identifying, and tracking key concepts and entities, leveraging external knowledge sources, and taking into account the social context of the dialogue. Effectively capturing and preserving the contextual meaning throughout the course of a conversation necessitates the utilization of various NLP techniques, including contextual embeddings, attention mechanisms, memory mechanisms, and discourse analysis [11]. By incorporating these techniques into the model architecture and training process, the model becomes more proficient at capturing the intricacies of language and generating responses that are more pertinent and coherent.

Figure 1 illustrates a general framework for a conversational agent, providing a fundamental overview of the key components typically found in such agents [12].

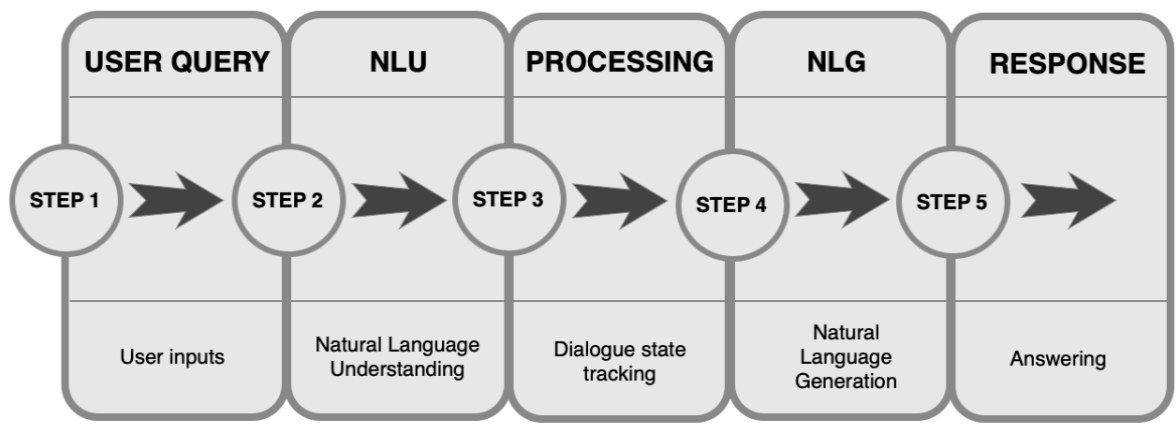

**Figure 1**: The general pipeline for conversational agent (adopted from [12])

Transformers [13] have demonstrated their efficiency in various natural language processing (NLP) applications, such as chatbots and conversational agents. Consequently, the challenge lies in two aspects: accurately identifying the user query within the dialogue and constructing a response based on the acquired data exploration outcomes.

## 4. Methods and Materials

We propose a research framework consisting of several key stages. Firstly, it is essential to define the domain in which collaborative analysis and BI can be effectively modeled. Given the diverse goals, preferences, experiences, and conditions of different users, they will access the same data with distinct requests, shaping the content of collaborative sessions. Secondly, it is crucial to identify and determine the data sources required for the analysis. Integrating data from various sources necessitates addressing challenges related to data consolidation, cleaning, and standardization. Thirdly, a fundamental stage involves constructing a knowledge base of collaborative decision-making cases. This entails developing an information model to capture data about each session, encompassing user behavior, research outcomes, and interactions with other users. Fourthly, it is imperative to design a user-friendly interface that facilitates data visualization and seamless interaction within the virtual space. The final stage revolves around processing, analyzing, and summarizing the collected data pertaining to user behavior. We firmly believe that through these efforts, we can establish a comprehensive CBI framework and validate models and technologies to support the virtual space, thereby enhancing the functionality of the BI4people project platform.

We commence our research by focusing on the task of data exploration and examining the potential benefits of utilizing a CBI virtual assistant to assist users in this process. In the proposed research it is suggested that the domain of study, relevant data sources, and consolidating the data are done. The general pipeline of preprocessing step is illustrated in Figure 2.

Load data involves acquiring and importing the relevant data from various sources into the CBI system. Once the data is loaded, the exploration phase begins. Users can analyze the data to gain insights, identify patterns, and understand the underlying trends. Metadata refers to the descriptive information about the data, such as its structure, format, and semantics. In this step, metadata is generated to provide a comprehensive understanding of the data. It includes attributes like data types, relationships, and relevant business rules, which facilitate data interpretation and usage. Users need to formulate queries to retrieve specific information or perform calculations on the data. Building queries involves constructing appropriate syntax and selecting relevant parameters based on the desired analysis or information requirements. Once the queries are formulated, they need to be matched with the appropriate commands or actions within the collaborative business intelligence system. This step ensures that the system understands and interprets the user's intent correctly. Generating prompts requires understanding the user's context, preferences, and the objectives of the analysis. The main goal of this step is transferring content to the NLG component. These steps collectively contribute to the preprocessing phase in CBI, enabling users to efficiently interact with the data.

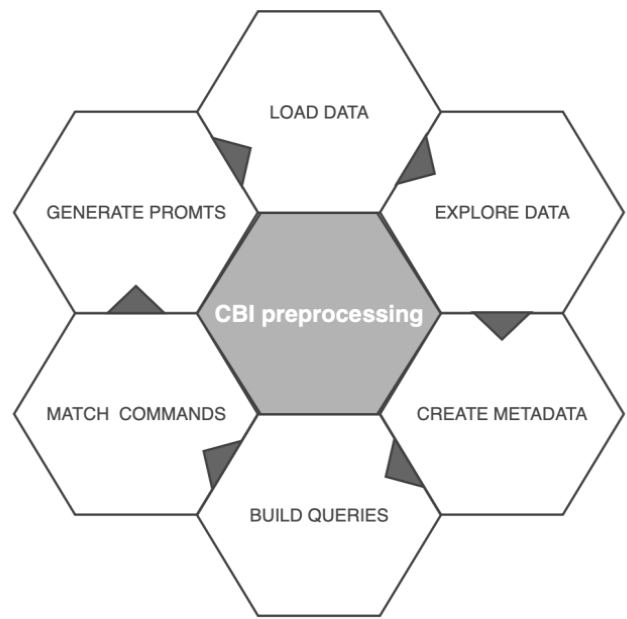

**Figure 2**: The general pipeline of preprocessing

Let us describe the data we use for experimenting. As a case-study we choose the traffic accidents in France. For each bodily accident occurring on a road open to public traffic, involving at least one vehicle and causing at least one victim requiring treatment, information describing the accident is entered by the law enforcement unit (police, gendarmerie) which intervened at the scene of the accident [14]. These entries are compiled in a form entitled bodily accident analysis report. All of these files constitute the national file of traffic accidents known as the "BAAC file" administered by the National Interministerial Road Safety Observatory "ONISR".

The databases, extracted from the BAAC file, list all the bodily injury accidents occurring during a specific year in mainland France and in the overseas departments with a simplified description. This includes accident location information, as entered, as well as information regarding the characteristics of the accident and its location, the vehicles involved and their victims. Every year, road accidents cause thousands of deaths. People are wondering what the causes are, what specific issues influence the most, who are under the risk etc. We use the dataset available at [14] as an example how people can explore data collaboratively and show how the Virtual Assistant can support them. The data consists of four datasets which are describe features of accidents (tabl. 1), places (tabl. 2), users (tabl. 3), and vehicles (tabl. 4).

**Table 1**
Data set "caracteristics.csv" [14]

| Title | Values | | | |
|---|---|---|---|---|
| *Feature* | *Num_Acc* | *an* | *mois* | *jour* |
| Description | Accident ID | Year of the accident | Month of the accident | Day of the accident |
| *Type* | *Int* | *Int* | *Int* | *Int* |
| Possible values | not specified | not specified | not specified | not specified |
| *Feature* | *hrmn* | *lum* | *agg* | *int* |
| Description | Time of the accident in hour and minutes | Lighting conditions in which the accident | Localization | Type of Intersection |
| *Type* | *Int* | *Int* | *Int* | *Int* |
| Possible values | not specified | 1 - Full day<br>2 - Twilight or dawn<br>3 - Night without public lighting<br>etc. | 1 - Out of agglomeration<br>2 - In built-up areas | 1 - Out of intersection<br>2 - Intersection X<br>3 - Intersection T<br>etc. |
| *Feature* | *atm* | *col* | *com* | *adr* |
| Description | Atmospheric conditions | Type of collision | Municipality | Postal address |
| *Type* | *Int* | *Int* | *Int* | *Str* |
| Possible values | 1 - Normal<br>2 - Light rain<br>3 - Heavy rain<br>4 - Snow - hail<br>5 - Fog - smoke<br>etc. | 1 - Two vehicles - frontal<br>2 - Two vehicles - from the rear<br>3 - Two vehicles - by the side<br>etc. | The commune number is a code given by INSEE. | variable filled in for accidents occurring in built-up areas |
| *Feature* | *gps* | *lat* | *long* | *dep* |
| Description | GPS coding | Latitude | Longitude | Department |
| *Type* | *Str* | *Int* | *Int* | *Int* |
| Possible values | M = Métropole<br>A = Antilles<br>G = Guyane<br>R = Réunion<br>Y = Mayotte | not specified | not specified | INSEE Code of the department followed by a 0 |

**Table 2**
Data set "places.csv" [14]

| Title | Values | | | |
|---|---|---|---|---|
| *Feature* | *Num_Acc* | *voie* | *V1* | *V2* |
| Description | Identifier of the accident | Route number | Numerical index of the road number | Alphanumeric index letter of the route |
| Type | Int | Int | Int | Str |
| Possible values | not specified | not specified | not specified | not specified |
| Feature | catr | circ | surf | plan |
| Description | Road category | Traffic regime | Surface condition | Plan layout |
| Type | Int | Int | Int | Int |
| Possible values | 1 – Motorway<br>2 – National road<br>…<br>9 – other way | -1 – Not specified<br>1 – One way<br>2 – Bidirectional<br>etc. | -1 – Not specified<br>1 – Normal<br>2 – Wet<br>etc. | -1 – Not filled in<br>1 – Straight part<br>2 – Curved left<br>3 – Curved right<br>4 – "S" shaped |
| Feature | vosp | prof | env1 | nbv |
| Description | the existence of a reserved lane, regardless of whether the accident took place on it. | Long profile describes the gradient of the road at the location of the accident | Maximum authorized speed at the place and at the time of the accident | Total number of traffic lanes |
| Type | Int | Int | Int | Int |
| Possible values | -1 – Not filled in<br>0 – Not applicable<br>1 – Cycle path<br>2 – Cycle lane<br>3 – Reserved lane | - 1 – Not specified<br>1 – Flat<br>2 – Slope<br>3 – Top of hill<br>4 – Bottom of hill | not specified | not specified |
| Feature | pr | pr1 | lartpc | situ |
| Description | Attachment PR number (upstream terminal number) | Distance in meters to the PR (compared to the upstream terminal) | Width of the central reservation (TPC) if it exists (in m) | Situation of the accident |
| Type | Int | Int | Int | Int |
| Possible values | The value -1 means that the PR is not filled in | The value -1 means that the PR is not filled in | not specified | -1 – Not specified<br>0 – None<br>1 – On the road<br>etc. |

The dataset contains a wide range of potential queries, each requiring the execution of appropriate commands or sequences of commands to filter, extract, process, and present the desired answer. By formulating specific queries, users can delve deeper into the data and gain valuable insights related to road safety and accident patterns. The users can inquire about various aspects of the data to explore it comprehensively. To gain general insights into accidents, users can pose the following questions:
- Is the number of accidents per year decreasing?
- Which months exhibit a higher frequency of accidents?
- Which day of the month is considered the safest for driving?

Moreover, users can leverage the data on road-specific accidents to inquire about the following:
- Which types of roads are associated with a high risk of accidents?
- What type of road gradient poses a high risk?

**Table 3**
Data set "users.csv" [14]

| Title | Values |
|---|---|

| Feature | Num_Acc | an_nais | num_veh | catu |
|---|---|---|---|---|
| Description | Accident ID | User's year of birth | Identification of the vehicle | User category |
| Type | Int | Int | Str | Int |
| Possible values | not specified | not specified | for each user occupying this vehicle - alphanumeric code | 1 – Driver<br>2 – Passenger<br>3 – Pedestrian |
| Feature | trajet | secu | locp | actp |
| Description | Reason for travel | the presence of safety equipment | Location of the pedestrian | Action of the pedestrian |
| Type | Int | Int | Int | Int |
| Possible values | -1 – Not specified<br>0 – Not filled in<br>1 – Home – work<br>2 – Home – school<br>3 – Shopping – purchases<br>…<br>9 – Other | -1 – Not filled in<br>0 – No equipment<br>1 – Belt<br>2 – Helmet<br>…<br>8 – Not determinable<br>9 – Other | -1 – Not filled in<br>0 – Not applicable<br>On pavement:<br>1 – A + 50 m from the pedestrian crossing<br>…<br>9 – Unknown | -1 – Not filled in<br>Moving<br>0 – Not filled in or not applicable<br>1 – Direction of vehicle hitting<br>…<br>9 – Other |
| Feature | etatp | sexe | place | |
| Description | whether the injured pedestrian was alone | User gender | Location of user | |
| Type | Int | Int | Int | |
| Possible values | -1 – Not filled in<br>1 – Alone<br>2 – Accompanied<br>3 – In a group | 1 – Male<br>2 – Feminine | the seat occupied in the vehicle (10 – Pedestrian) | |

In the context of data exploration, various aspects related to the people involved in accidents can be investigated. This includes analyzing the condition of individuals after the accident, examining the age distribution of those involved, and exploring the distribution of sexes among the individuals affected. Additionally, data exploration can also focus on the use of safety equipment, such as investigating the distribution of safety equipment usage and determining whether the use of safety equipment had any impact on the condition of people after the accident.

The data exploration process implemented on the accident dataset can involve a comprehensive analysis of various factors and dimensions related to the accidents. This include examining the characteristics and demographics of the people involved, such as their age, sex, and condition after the accidents. Additionally, the exploration encompasses investigating the use of safety equipment and its potential impact on the outcomes of the accidents. By delving into these aspects, the data exploration aims to uncover patterns, trends, and insights that could contribute to a better understanding of the accident occurrences and inform strategies for improving safety measures.

Our research focuses on enhancing data exploration by investigating how a CBI virtual assistant can effectively identify user tasks and transform them into executable commands. The primary objective is to enable the virtual assistant to understand user intentions and translate them into specific actions that can be executed within the data exploration process. By developing robust techniques for task identification and command generation, we aim to enhance the usability and efficiency of the virtual assistant, enabling users to seamlessly interact with the data and extract meaningful insights.

**Table 4**
Data set "vehicles.csv" [14]

| Title | Values | | | |
|---|---|---|---|---|
| Feature | Num_Acc | senc | occutc | num_veh |
| Description | Accident ID | Flow direction | Number of occupants in public transport | Identification of the vehicle |
| Type | Int | Int | Int | Str |
| Possible values | not specified | -1 – Not filled in<br>0 – Unknown<br>1 – PK or PR<br>…<br>3 – No mark | not specified | for each user occupying this vehicle - alphanumeric code |
| Feature | catv | obs | obsm | choc |
| Description | Category of vehicle | Fixed obstacle struck | Moving obstacle struck | Initial shock point |
| Type | Int | Int | Int | Int |
| Possible values | 00 – Indeterminable<br>01 – Bicycle<br>02 – Moped<br>03 – Cart<br>…<br>38 – Bus<br>…<br>99 – Other vehicle | -1 – Not specified<br>0 – Not applicable<br>1 – Parked vehicle<br>…<br>16 – Obstacle-free road exit<br>17 – Nozzle – aqueduct head | -1 – Not specified<br>0 – None<br>1 – Pedestrian<br>2 – Vehicle<br>4 – Rail vehicle<br>5 – Domestic animal<br>6 – Wild animal<br>9 – Other | -1 – Not specified<br>0 – None<br>1 – Before<br>2 – Front right<br>3 – Front left<br>4 – Back<br>…<br>9 – Multiple shocks (barrels) |

In summary, the research methodology comprises several significant steps. Firstly, the preprocessing of datasets plays a crucial role, involving activities such as data cleaning, exploration, and annotation. An essential aspect of this step is the creation of metadata descriptions, which aid in matching user queries to appropriate commands and facilitating scenario execution. Secondly, it is essential to establish a set of predefined basic commands for data exploration. This necessitates the development of a tool capable of generating abstract semantic trees for each piece of data and command, enabling the creation, or matching of executed code. Lastly, a conversational unit is required to capture user queries and generate responses based on the established scenario. This unit acts as the bridge between the user and the data exploration process, facilitating effective communication and interaction.

## 5. Results

To demonstrate the effectiveness of incorporating a conversational agent for data exploration, we present a scenario involving the analysis of road accidents in France [14]. In this scenario, our virtual assistant is equipped with a collection of statistical commands derived from the Scipy and Sklearn libraries, as well as text analysis capabilities.

The scenario commences with the loading of the data into a data frame. In our experiments, we perform data loading, cleaning, and labeling procedures in advance, ensuring the dataset is prepared for analysis.

To illustrate how we can utilize data and the support provided by a CBI virtual assistant, let us consider an example. Prior to beginning, we load the data and preprocess it to ensure its completeness and suitability for analysis. In this particular scenario, we aim to answer the query, "What weather conditions are associated with the most accidents?" To initiate a conversation regarding weather conditions, we match the topic with the relevant data feature, namely "atm" (atmospheric conditions). As a first step, we clean the data by removing any accidents that have a missing value for "atm." We identified and removed 55 accidents with NA values. Subsequently, we can proceed to analyze the distribution of accidents across different weather conditions, as depicted in Figure 3, to determine which weather condition is associated with the highest number of accidents.

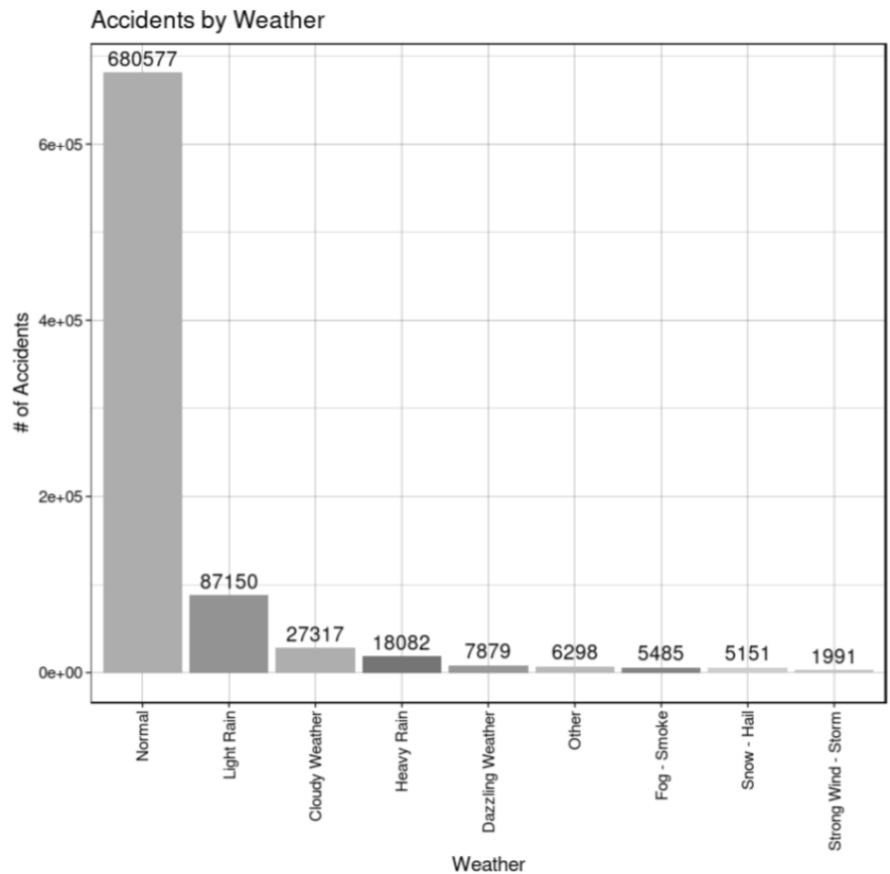

**Figure 3**: The number of accidents over different weather conditions

To find the answer we run the code depicted on the figure 4. Firstly, we define the function which calculate the number of unique values in the column. The function returns the maximum value. We predefine that dataset describes accidents where each row presents an accident. So, the function can be used to calculate the most accidents according to any possible query. For example, we can find what year, month, lighting conditions, type of intersection or municipality.

```python
def most_of_accidents (x,y):
    most_of_accidents = x[y].value_counts().idxmax()
    return most_of_accidents
most_of_accidents (weather_data,'atm')
```

**Figure 4**: Python code of most_of_accidets() function

Then we need to explain the function in order to compare it with user's query. We suggest building the abstract semantic tree as it is depicted on the fig.5. Follow the semantic tree we can find that there are two variables and two methods which are called. But it is not enough in terms of user's query identification. Also, we need to prepare and describe metadata of the dataset as well as create the trace table between semantic of command and parts of abstract tree. It gives us a tool for matching user's query and command to choose appropriate one.

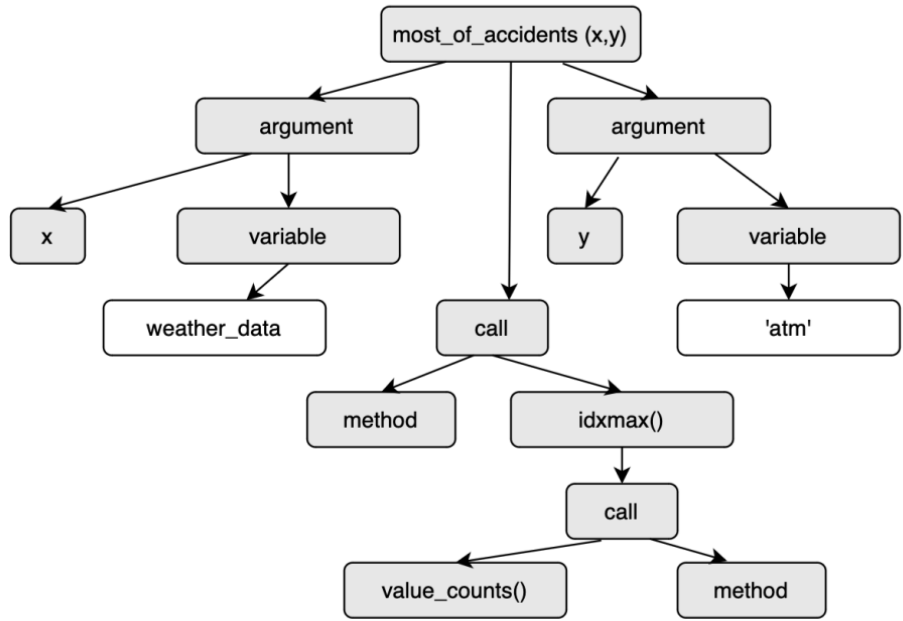

**Figure 5**: Abstract semantic tree

The example of dialogue is presented on the fig.6. The virtual assistant has to parse the expression "What weather has the most accidents?" and identify the command, the attribute, and conditions. The trickiest is to match the word "weather" and the title of column "atm". In order to solve the task, we use meta description which contains explanation all of the attributes as well as synonyms and matching word. This step should be done in advance.

As evidenced by analysis, state-of-the-art linguistic models enable effective dialogue construction. These models require less training data and demonstrate increased robustness to errors, allowing for accurate classification and execution of each conversational step. The same principle is proposed for the implementation of a CBI virtual assistant. Through interactions with the user, the agent should be able to identify and classify commands, and if necessary, clarify parameters associated with them.

Based on the conducted analysis, it is evident that the integration of several crucial features within virtual assistant software is essential for effective user assistance. These features encompass:

- Versatile Data Exploration Commands: The virtual assistant must possess the capability to execute a diverse range of data exploration commands, including filtering, querying, selecting, and parameter configuration.
- Personalization based on User Preferences: The virtual assistant should incorporate personalization mechanisms that consider user preferences, historical interactions, and behavioral patterns. This personalized approach enhances the relevance and effectiveness of the recommendations provided by the assistant.
- Machine Learning Algorithms: The inclusion of machine learning algorithms enables the virtual assistant to continuously learn from user interactions, thereby improving its ability to deliver personalized recommendations over time.
- Language Understanding and Text Generation: Proficiency in language understanding and text generation is imperative for effective communication between the virtual assistant and the user. This capability enables the assistant to comprehend user queries and generate appropriate responses in a manner that promotes clear and meaningful interaction.

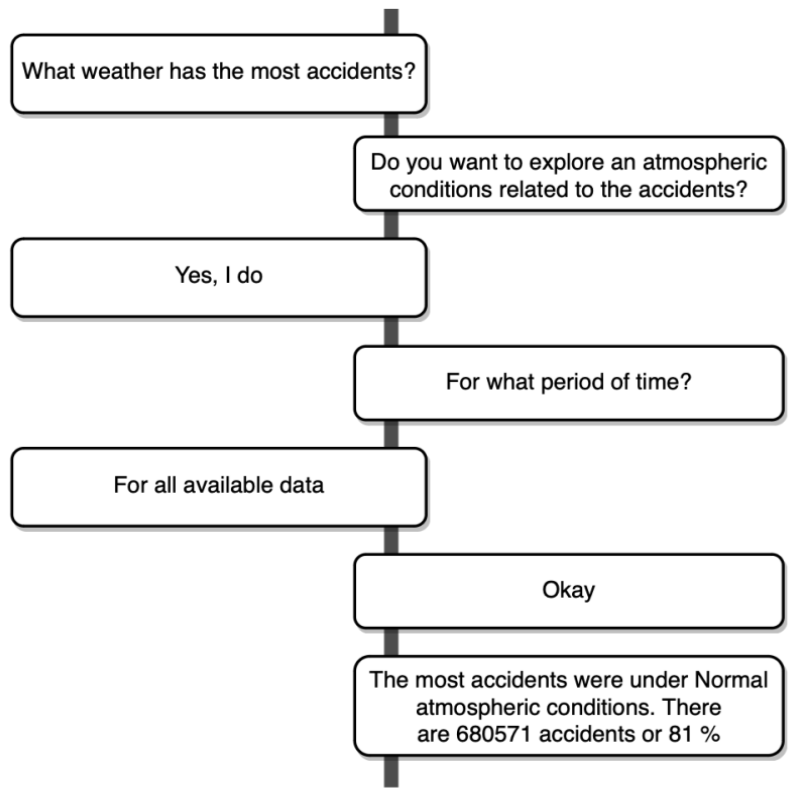

**Figure 6**: Dialogue with Virtual assistant example

By incorporating these essential features, virtual assistant software can optimize user assistance, enabling more efficient and effective support in various domains.

## 6. Discussion and Conclusion

In the rapidly evolving business landscape, there is a growing need for innovative methodologies that integrate intelligent technologies and tools to provide accurate and reliable information for effective decision-making. This paper presents an approach aimed at creating, validating, and evaluating collaborative decision-making models. Business Intelligence systems are extensively used to support decision-making processes in various organizations. Collaborative Business Intelligence offers even greater opportunities for informed decision-making by leveraging external information from diverse sources.

The proposed CBI virtual assistant aims to enhance the accessibility of data exploration for a wider range of users while reducing the time and effort required for data analysis. The key contributions of this research include: 1) a reference model representing collaborative BI, inspired by linguistic theory; 2) an approach to transform user queries into executable commands, enabling their integration into data exploration software; and 3) the primary workflow of a conversational agent designed for data analytics. The next phase of this research involves implementing a prototype of the CBI virtual assistant and conducting experiments to evaluate its effectiveness.

## 7. Acknowledgements

The research study depicted in this paper is funded by the French National Research Agency (ANR), project ANR-19-CE23-0005 BI4people (Business intelligence for the people).